\documentclass[12pt,preprint]{aastex63}
\usepackage{graphicx}
\usepackage[]{xcolor}

\received{...}
\revised{...}
\accepted{...}
\submitjournal{ApJ}

\begin{document}

\title{The Bulk Flow Velocity and Acceleration of the Inner Jet in M\,87}

\correspondingauthor{Brian Punsly}
\email{brian.punsly@cox.net}

\author{Brian Punsly}
\affiliation{1415 Granvia Altamira, Palos Verdes Estates CA, USA 90274}
\affiliation{ICRANet, Piazza della Repubblica 10 Pescara 65100, Italy}
\affiliation{ICRA, Physics Department, University La Sapienza, Roma, Italy}

\begin{abstract}
A high sensitivity, 7mm Very Long Baseline Array image of M\,87 is analyzed in order to estimate the jet velocity within 0.65 mas of the point of origin. The image captured a high signal to noise, double-ridged, counter-jet extending $\sim 1$ mas from the nucleus. After defining conditions and requirements that justify approximate time averaged bilateral symmetry, a continuous set of Lorentz transformations are found that map the double-ridged counter-jet intensity profile into the double-ridged jet intensity profile. The mapping is realized by a uniformly accelerating flow with intrinsic velocity of $\sim 0.27$c at 0.4 mas (a de-projected distance of 0.38 lt-yrs) to $0.38$c at 0.65 mas (a de-projected distance of 0.61 lt-yrs) from the nucleus. Since the velocity field is derived from the global surface brightness profile and does not depend on the motion of enhanced features, it is most likely a bulk flow velocity as opposed to a pattern velocity. This interpretation is corroborated by the fact that the distribution of the apparent velocities of previously identified individual features (from the literature) within 0.65 mas of the nucleus are consistent with local hydrodynamic shocks being advected with the local bulk flow velocity. The bulk flow velocity of the visible inner jet is a constraint that can potentially break degeneracies between numerical simulations that are designed to replicate both the annulus that was imaged by the Event Horizon Telescope as well as the base of the inner jet.
\end{abstract}


\section{Introduction}

The galaxy, M\,87, is the host of the nearest powerful extragalactic radio source, 3C\,274, and is a prime target of the Event Horizon Telescope (EHT).
Even with an eventual EHT image of the jet base, lower frequency Very Long Baseline Interferometry (VLBI) {is required in order to ascertain the physical nature of the jet. Due to gravitational lensing and redshifting by the central black hole, the jet launching region will probably not be seen directly. The EHT image of M\,87 is dominated by a thin annulus comprised of direct emission and a bright lensing (or photon) ring that is the focus of emission from the back side of the disk and counter-jet at $r\approx5 M$, where $M \approx 6 \times 10^{9} M_{\odot} \approx 8.8 \times 10^{14} \rm{cm}$ is the black hole mass in geometrized units \citep{bar73,aki19,aki21,gra19}. Extricating the direct jet flux has no unique decomposition. Furthermore, any of the so far undetected, direct jet flux originating close the horizon is highly redshifted. Without seeing the jet in future EHT images extend from larger distances down to its launch point, determination of the point of origin cannot be achieved directly. One must rely on assumptions and models to argue for a particular point of origin. Thus, the basic properties of the jet adjacent to this region are crucial for determining which jet launching model and assumptions recreate the physics of the jet. For example, the jet opening angle is an important constraint. High resolution VLBI observations of M\,87 indicate an extremely edge brightened jet with an extremely wide jet opening angle within 0.1\,mas of the
core, $\sim 127^{\circ}$  \citep{had16,kim18,pun19}. Similarly, 3C\,84 (another bright nearby extragalactic jet), has a very edge brightened jet within 0.1\,mas of the
core, and an opening angle of $138^{\circ}$, the maximum possible angle, geometrically, is $180^{\circ}$ \citep{gio18}. This wide opening angle and extreme edge brightening might be characteristic of all low luminosity active galactic nuclei with powerful jets. It has been previously shown that the large opening angle and the large degree of edge brightening in the innermost jet of M\,87 are a challenge for current numerical models of jet launching \citep{pun19,cha19,mos16}. In this study, another property on the inner jet is investigated that will also constrain any numerical model that attempts to simulate both the EHT detected annular emission and the inner jet, the bulk velocity of the jet close to the nucleus.
\par Historically, the jet velocity near the source has been ambiguous to determine. One usually finds component motion in VLBI images, but it is unclear if this is bulk flow speed or a surface pattern speed \citep{lis09,jor17}. Furthermore, it is an apparent speed that is affected by Doppler abberation \citep{ree66}. In this paper, a remarkable high sensitivity image from the 43 GHz Very Long Baseline Array (VLBA) is used to carefully estimate the jet speed from the Doppler enhancement of the approaching jet relative to the receding jet. Section 2 is a discussion of the image and the double-ridged intensity profiles of the jet and counter-jet. The next section is a critical analysis of the circumstances for which bilateral symmetry is an appropriate approximation for a jetted system. Section 4 describes how the counter-jet and jet profiles are related using special relativity if the requirements of the bilateral symmetry approximation are met. The findings are discussed in the context of other results in the literature in Section 5. Throughout the paper, a line of sight to the jet (LOS) of $18^{\circ}$ is assumed \citep{cha19}. The
following cosmological parameters are dopted: $H_{0}$=69.6 km s$^{-1}$Mpc$^{-1}$, $\Omega_{\Lambda}=0.714$ and $\Omega_{m}=0.286$ and use Ned Wright’s Javascript
Cosmology Calculator \footnote{http://www.astro.ucla.edu/~wright/CosmoCalc.html} \citep{wri06}.

\section{Intensity Profiles of the High Sensitivity Image}
The VLBA image from January 12, 2013 was observed in the wide-band mode which increased sensitivity. The bandwidth was 256 MHz in each of 2 polarizations for 512 MHz total.  The sample rate is 2 Gbps for Nyquist sampling ($2\times$ bandwidth) and 2 bits per sample. Previous observations had been at 512 Mbps in 2011 and 2012 and 256 Mbps for most epochs before that. Details are in Table 3 of \citet{wal18}. The image also has superior coverage in the $(u,v)$ plane compared to the other wide-band observations (R. Craig Walker private communication 2021). Mild super-resolution of 70\% of the primarily north-south major axis of the Briggs weighted beam was used to restore the data in the previously published image \citep{briggs95,wal18}. The resulting image clearly shows a $\sim 1$mas long double-ridged counter-jet. This feature is what makes this image particularly useful for this study. Using the image FITS file that was generously provided by R. Craig Walker, the image is re-plotted in Figure 1. The convolving beam is 0.21 x 0.16 mas at PA = $0^{\circ}$. The Briggs weighted beam is 0.31 x 0.16 mas at PA =$ -5.28^{\circ}$. The image is rotated so that the jet axis is horizontal with an axial coordinate, $z$.
\begin{figure}
\begin{center}
\includegraphics[width=170mm, angle=0]{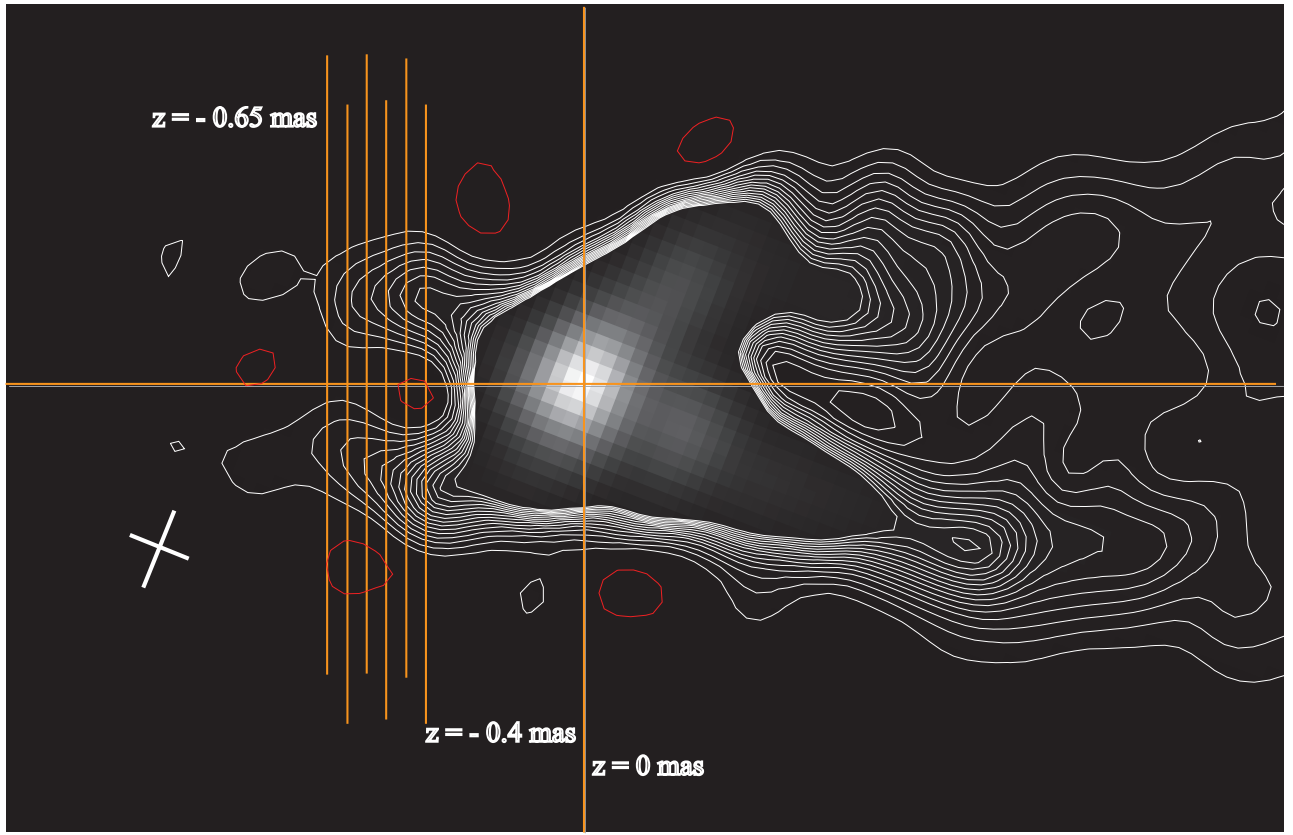}
\caption{A high sensitivity 43\,GHz VLBA image from January 12, 2013 that is designed to highlight the counter-jet. The jet is rotated so the horizontal axis aligns with the jet. The north-south major axis of the restoring beam (the cross in the lower left) therefore appears tilted. A z-axis with a grid is shown. The white linear contours levels are (1, 1.64, 2.29 ..... 10) mJy/beam. The red contours are at -0.5 mJy/beam. The image is a tool for estimating the peak ridge surface brightness of the counter jet in six, 0.05 mas, increments from -0.4 mas to -0.65 mas as indicated by the orange lines.}
\end{center}
\end{figure}
The surface brightness is concentrated on the two ridges that are largely unresolved in the north-south direction \citep{had13,had16}. Thus, in order to compare the brightness ratios of the jet and counter jet, it is useful to estimate the peak surface brightness of the two ridges as a surrogate for the flux density of a cross-section since these are the high signal to noise, distinct features. To facilitate this effort, the high sensitivity image in Figure 1 was created. The main difference from the published image is the high resolution linear scale (indicated by the white contours), that goes from 1 mJy/beam to 10 mJy/beam in 15 increments. The red contours are at -0.5 mJy/beam. The computational advantage is that the peak intensity along each ridge can be read right off the contour map with $< 0.3$ mJy/beam accuracy. Figure 1 highlights the faint counter-jet at the expense of having the nuclear region saturated. Two other images with higher values of the linear contour levels were also made to capture the jet, with linear scales of 10-25 mJy and 25-50 mJy. The core surface brightness extends to $\mid z\mid\sim$ 0.3-0.4 mas \citep{wal18}. Thus, the following analysis is restricted to $\mid z\mid \geq 0.4$ mas. The peak intensities, $I_{N}(z)$ and $I_{S}(z)$, of the north and south ridges, respectively, estimated from these images are listed in columns (2) and (3) of Table 1.
\begin{table}
\begin{center}
 \caption{Peak Intensity and Flux Density of Jet and Counter-Jet}
\begin{tabular}{cccccc}
 \hline
 $z$  & $I_{N}(z)$  & $I_{S}(z)$ &  $I(z)\equiv I_{N}(z) + I_{S}(z)$ & Flux Density \\
 &  &  &  & of 0.16 mas wide \\
 &  &  &  & cross-section \\
  (mas)   & (mJy/beam)  & (mJy/beam) &  (mJy/beam)& (mJy) \\
 \hline
 -0.65 & $1.4 \pm 0.9$  & $1.9 \pm0.9$ & $3.3 \pm 1.0$ & $3.8\pm 1.0$ \\
-0.60 & $2.1 \pm 1.0$  & $2.4 \pm 1.0$ & $4.5 \pm 1.0$& $5.6 \pm 1.1$ \\
 -0.55 & $2.8 \pm 1.0$  & $3.1 \pm 1.0$ & $5.9 \pm 1.1$& $6.1 \pm 1.1$\\
 -0.50 & $3.4 \pm 1.0$  & $3.9 \pm 1.0$ & $7.3 \pm 1.2$& $7.8 \pm 1.2$\\
-0.45 & $4.4 \pm 1.0$  & $5.5 \pm 1.1$ & $9.9 \pm 1.4$ & $12.1\pm 1.5$\\
-0.405\tablenotemark{a}&     &   & & $14.9 \pm 1.8$\\
 -0.40& $5.5 \pm 1.1$  & $7.2 \pm 1.2$ & $12.7 \pm 1.6$& $14.5 \pm 1.7$\\
 0.4 & $24.5 \pm 2.8$  & $35.0 \pm 3.8$ & $59.5 \pm 6.4$ & $74.0 \pm 7.5$ \\
0.405\tablenotemark{a}&     &   & & $66.4 \pm 6.7$\\
0.45 & $19.1 \pm 2.1$  & $28.5 \pm 3.0$ & $47.6\pm 4.9$& $52.2 \pm 5.3$\\
 0.5 & $14.3 \pm 1.7$  & $25.9 \pm 2.8$ & $40.2 \pm 4.1$& $44.6 \pm 4.6$ \\
 0.55 &  $12.1 \pm 1.5$  & $21.3 \pm 2.3$ & $33.4 \pm 3.5$ & $33.5 \pm 3.5$\\
0.6 & $11.7 \pm 1.5$  & $18.9 \pm 2.1$ & $30.6 \pm 3.2$& $32.2 \pm 3.4$ \\
0.65 & $11.3 \pm 1.5$  & $16.4 \pm 1.9$ & $27.7 \pm 2.9$& $33.2 \pm 3.4$
\end{tabular}
\end{center}
\tablenotetext{a}{A small shift in the innermost point removes a strong CC from the 0.16 mas window (it is only 0.323 mas from core) and lowers the flux density/intensity ratio.}
\end{table}
We also use the CLEAN component (CC) model associated with this image (generously provided by R. Craig Walker) in order to estimate the flux density of cross sections of the jet. A cross-sectional width of 0.16 mas, equal to the east-west full width at half maximum of the Briggs weighted beam (the jet is predominantly east-west), is chosen. These are listed in the last column of Table 1. The CC model is not a perfect reconstruction of M\,87, but is an alternative to the ridge intensities to study the jet surface brightness that does capture flux between the ridges. It provides a consistency check, since the two diagnostics should track each other. Indeed, they track each other except at $z= +0.4$ mas (this outlier is discussed in detail in Section 4).
\par Two factors are considered in determining the uncertainties in Table 1. An uncertainty in the intensity of distinct features in 22 GHz VLBA images was estimated at 5\% - 10\% \citep{hom02}. 43 GHz observations are more difficult, but this is an exceptionally good image, so an uncertainty of 10\% in all measured feature intensities is assumed. For the very faint features, such as the counter-jet, the noise level is the dominant source of uncertainty. Inspection of the negative contours in the image near the counter-jet reveals numerous adjacent contours between -0.25 mJy/beam and -0.93 mJy/beam. Thus, a background noise uncertainty of 0.93 mJy/beam is added to the 10\% in quadrature in Table 1. For this reason, the counter-jet analysis is cutoff at 0.65 mas. At 0.7 mas the intensity is comparable to the magnitude of the negative contours.
\begin{figure*}[htp!]
\begin{center}
\includegraphics[width=120mm, angle=0]{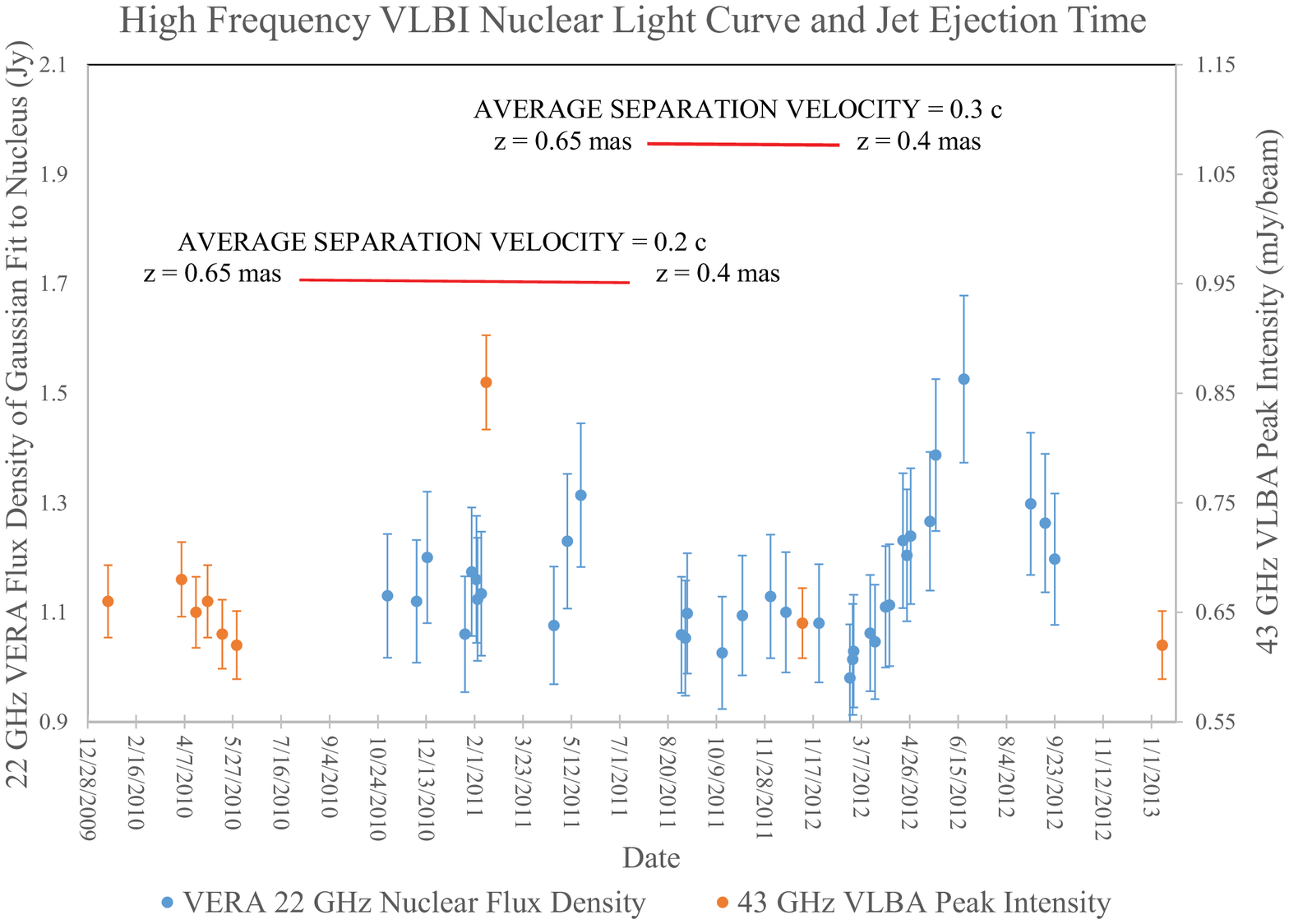}
\includegraphics[width=120mm, angle=0]{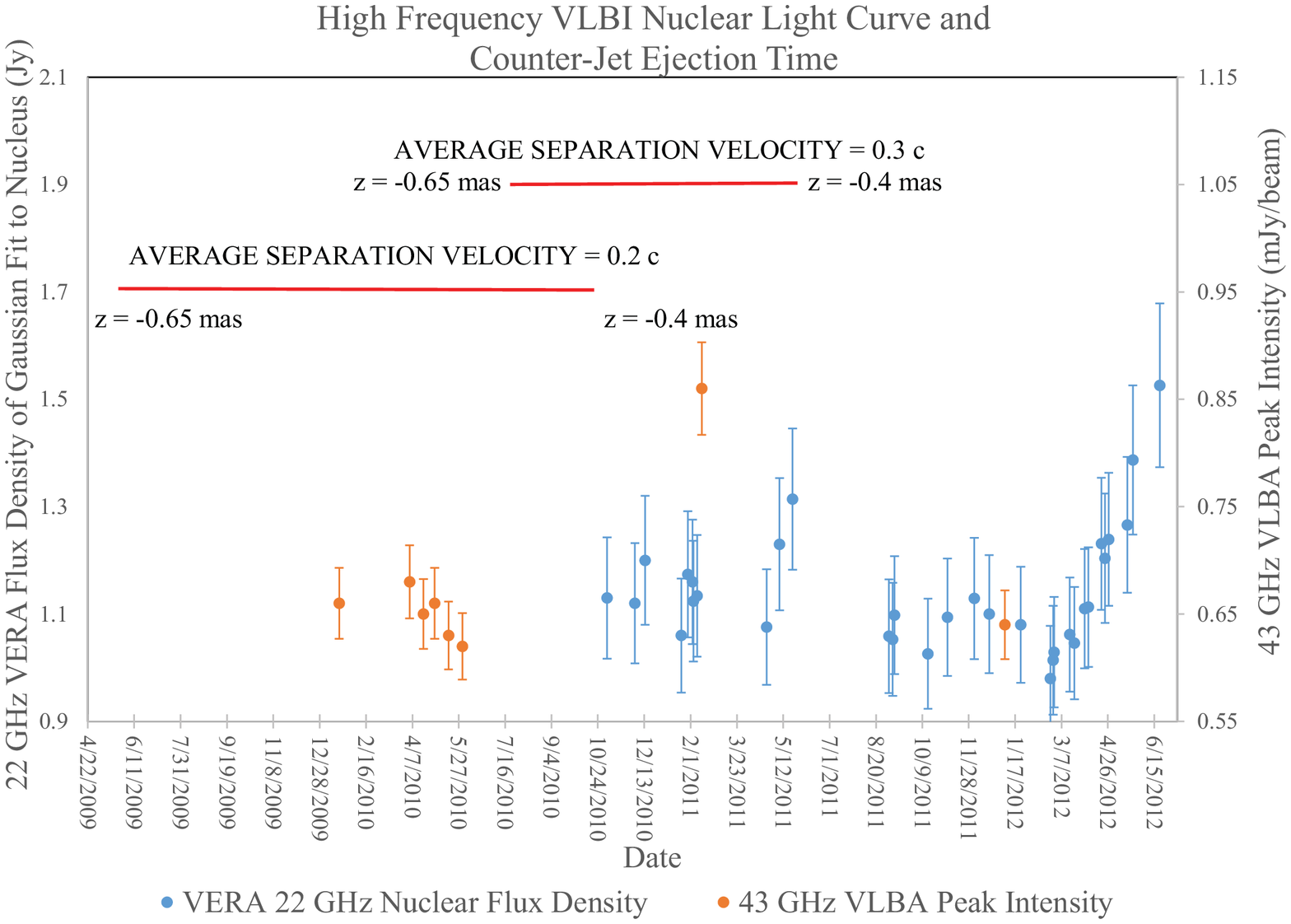}
\caption{The figure is used investigate nuclear activity during plausible ejection times that depend on the average jet velocity during the time of flight from the point of origin. We have surprisingly good coverage of the nuclear region during the plausible time frames except for the counter-jet, in the bottom frame, between June and December 2009.}
\end{center}
\end{figure*}

\section{Time Dilation and Bilateral Symmetry}
This analysis assumes approximate intrinsic bilateral symmetry of the jet and counter-jet in a time averaged sense. For example, in 3-D numerical simulations of accretion onto black holes, the jet has episodes of predominantly one-sided ejection, but on average the jet power is the same in each hemisphere \citep{pun09,mck12}.  There are two effects to consider in order to assess the possibility of observing bilateral symmetry, time dilation and the environment in which the jet propagates.
\subsection{Non-Simultaneous Ejections}
Firstly, for a given global (or observer) time, $t$, the receding jet will not have appeared to move as far as the approaching jet since the source of light is moving away from the observer and it takes longer for the signal to arrive. Estimates of a constant velocity based on this notion include the arm length ratio of double radio sources, \citet{sch95}, and bilateral ejections in Galactic X-ray binaries \citep{mir94,fen99}. The distance, $L$, traversed in the counter-jet direction and the jet direction are given by \citep{ree66,gin69}:
\begin{equation}
\frac{d\,L_{\rm{counter-jet}}(t)}{d\,t} =\frac{\beta_{-}(z)\sin{\theta}}{1-\beta_{-}(z)\cos{\theta}} \;, \rm{and}\; \frac{d\, L_{\rm{jet}}(t)}{d\,t}=\frac{\beta_{+}(z)\sin{\theta}}{1-\beta_{+}(z)\cos{\theta}}\;,
\end{equation}
where $\beta_{-}(z)$ is the 3-velocity (divided by c) of the receding jet and $\beta_{+}(z)$ is the 3-velocity of the approaching jet measured in the cosmological rest frame of M\,87 (with $\theta = 90^{\circ}$) and $\theta$ is the LOS. For approximate, time averaged bilateral symmetry (that is also averaged over the significant volume sampled by the synthesized beam of the VLBA), one might expect that
\begin{equation}
\beta_{-}(-z) \approx - \beta_{+}(z)\;.
\end{equation}
As an example, for a constant $\beta_{+}(z)$ and Equation (2), the arm length ratio, $R$, from Equation (1) is \citep{sch95},
\begin{equation}
R= \frac{1+\beta_{+}(z)\cos{\theta}}{1-\beta_{+}(z)\cos{\theta}}\;.
\end{equation}
For $\beta_{+}(z)=0.3$c and Equations (2) and (3), plasma at z = -0.5 mas in the counter jet was emitted at the same time, $t$, as plasma at z = +0.9 mas in the jet. Thus, comparing the jet to the counter jet is nontrivial.
\par In order to assess the potentially deleterious effects of time dilation to the analysis, one needs to know if the nucleus of M\,87 was approximately steady during the long time frame over which both the regions, $-0.65 \rm{mas} < z < -0.4\rm{mas}$ and $0.4 \rm{mas} < z < 0.65\rm{mas}$, were emitted. This analysis is attempted in Figure 2. We combine the peak intensities (the core) from 43 GHz VLBA with the Gaussian fits to the nucleus derived from 22 GHz VLBI Exploration Radio
Astrometry (VERA) observations in a light curve that fortuitously covers most of the relevant time frame \citep{wal18,had12,had14}. Estimating the time of ejection is equivalent to estimating the average velocity since the ejection. The values of 0.2c and 0.3c used in Figure 2 cover the plausible range that is consistent with the jet velocities found in the next section. It seems that, for the most part, M\,87 is in a low state except for one or two short flares. Thus, it is concluded that the attempted analysis is plausible if many jet cross-sections are considered in consort because there is a chance that some of the jet was ejected during modest flares. In summary, the relatively steady nature of the nucleus implies that comparing $I(z)$ to $I(-z)$, $0.4 \rm{mas} < z < 0.65\rm{mas}$, on January 12, 2013 has a reasonable chance to yield meaningful results.

\subsection{Propagation in an Inhomogeneous Environment} Secondly, the jet interacts with the environment and therefore the dissipation might be different in each hemisphere \citep{sch95}. Thus, one cannot compare the intensity at one point in the jet to the intensity at one point in the counter-jet. Formally, bilateral symmetry is (x, y, z)$\rightarrow$(x, y, -z). But clearly, this is only valid in a time averaged sense. However, due to environmental factors, the jet dissipation and therefore its brightness is an unknown function of the location in the (x, y) plane at any given time. One must also include averaging over the (x, y) plane in addition to time averaging in order to expect the bilateral symmetry approximation to be justified. Even so, it is prudent to compare ensembles of cross-sections of the jet at +z, averaged or integrated over the (x, y) plane, to cross-sections of the counter-jet at -z, averaged or integrated over the (x, y) plane, in order to mitigate the effects of the environment and non-simultaneous ejection times. This is the strategy of the calculation presented in the next section. An exact parity symmetry (x, y, z)$\rightarrow$(-x, -y, -z) was assumed in a previous jet/counter-jet brightness profile analysis of this image \citep{wal18}. This is not advocated here based on the discussion above.

\begin{figure}[htp!]
\begin{center}
\includegraphics[width=125mm, angle=0]{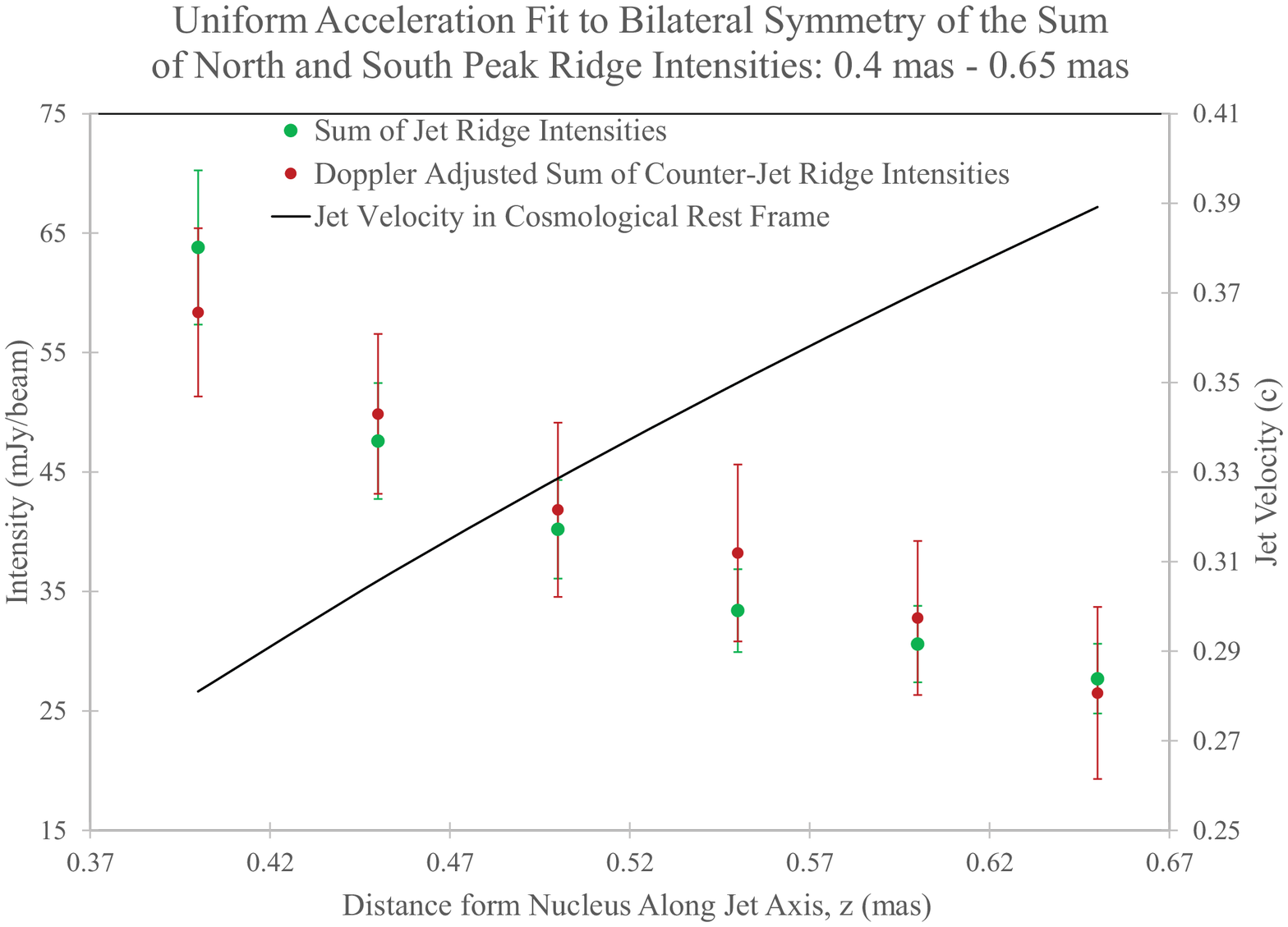}
\includegraphics[width=85mm, angle=0]{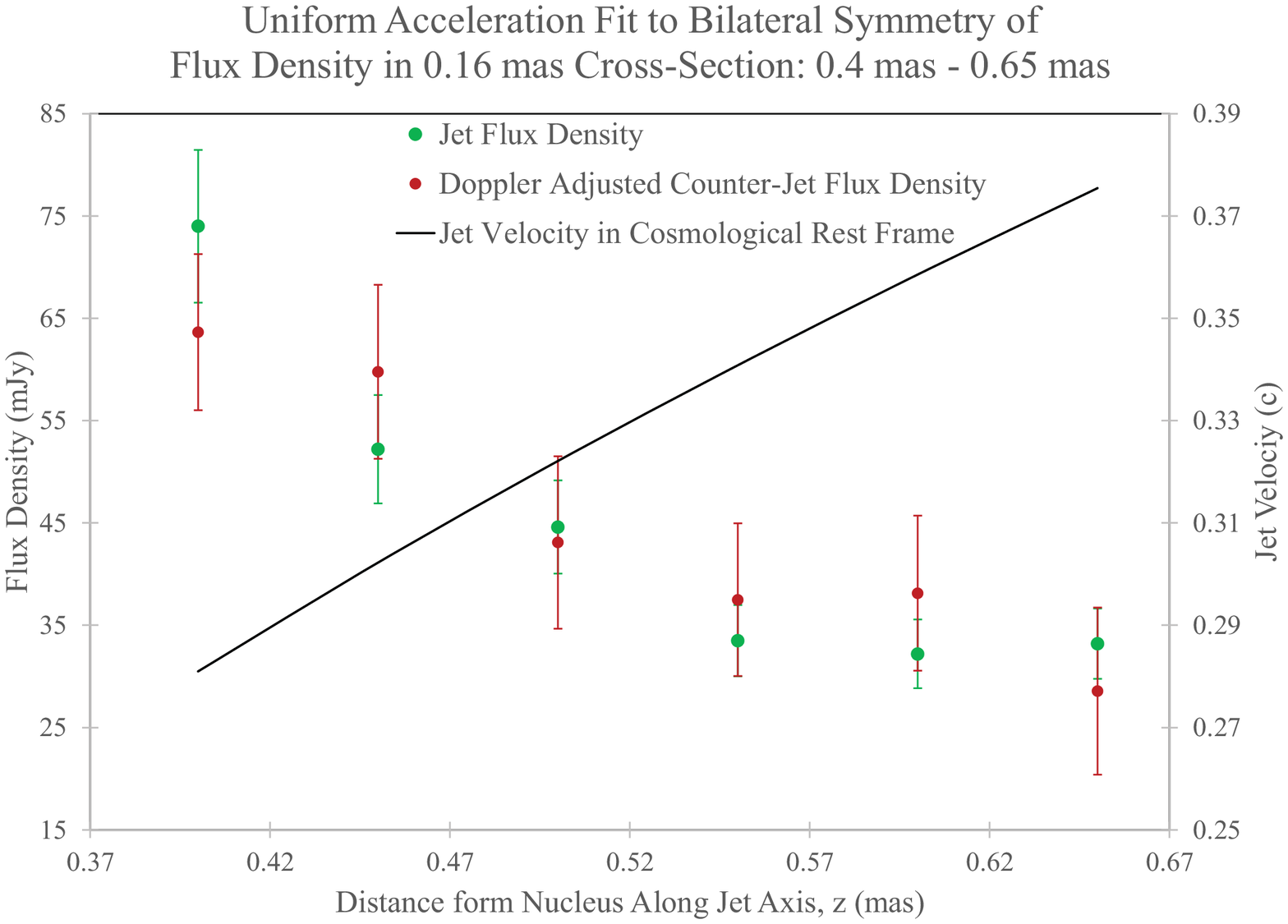}
\includegraphics[width=85mm, angle=0]{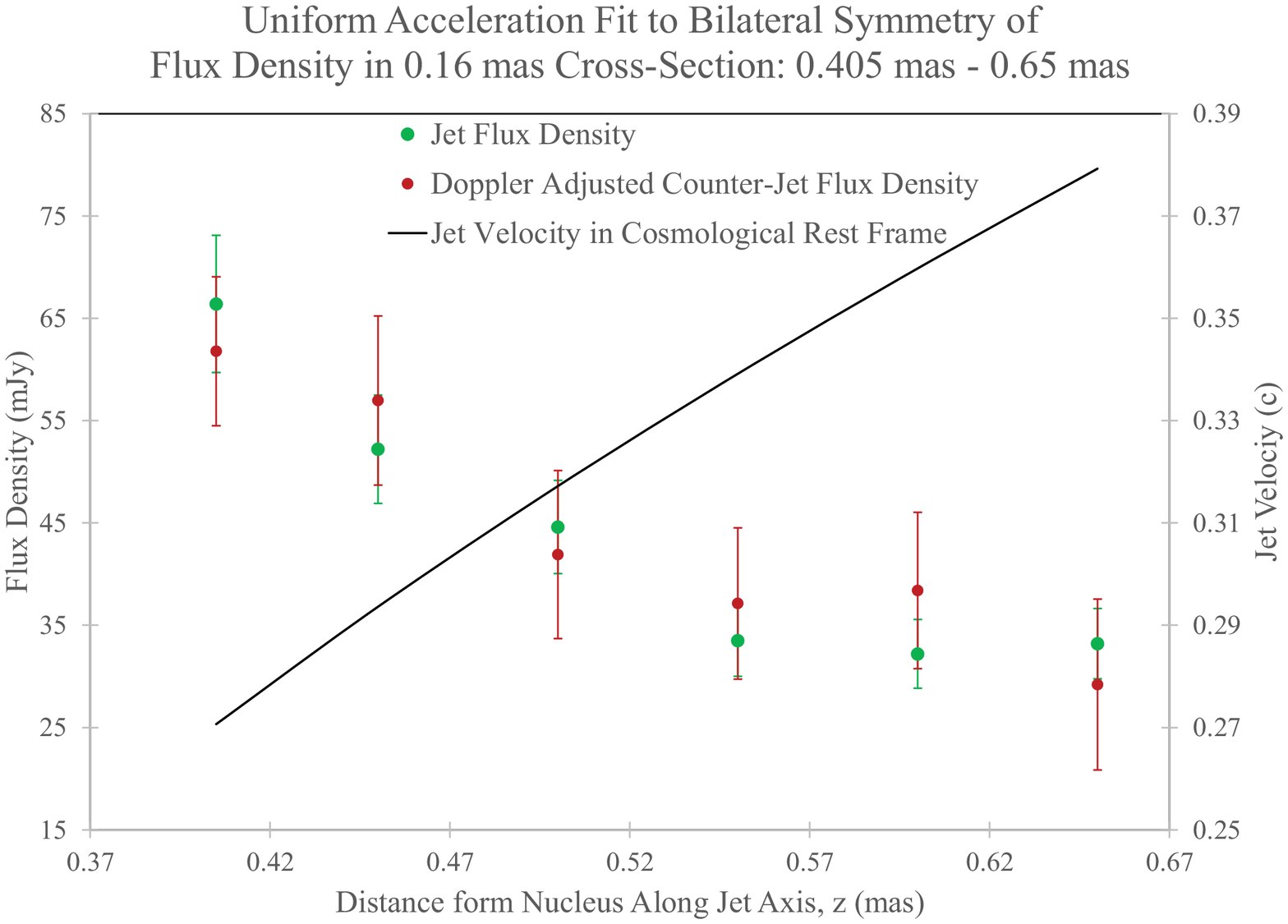}
\caption{The top panel is the fit to the sum of the ridge peak intensities based on Equations (2) - (4) and the jet velocity profile (the black curve in the figure). The Doppler adjusted counter-jet intensity is $\{R[\beta(z)]\}^{2.7}I(-z)$. The bottom frames are two fits for the flux density of a 0.16 mas wide cross-sections for the jet and counter-jet. The fit on the right is considerably better by just shifting the innermost point 0.005 mas from the bright nucleus.}
\end{center}
\end{figure}

\section{Relativistic Transformation from Counter-Jet to Jet}

The spectral intensity is assumed to be an approximate power law in frequency, $\nu$, $I_{\nu}\propto \nu^{-\alpha}$. Assuming perfect bilateral symmetry \citep{mir94,fen99},
\begin{equation}
I_{\nu}(z) \approx R^{k + \alpha} I_{\nu}(-z) \;, \quad z > 0\;.
\end{equation}
In Equation (4), $k=2$ for a resolved jet \citep{lin85}.  In the inner 1 mas of the jet, it was found that $\alpha \approx 0.6$ from 22 GHz to 43 GHz and $\alpha \approx 0.8$ from 43 GHz to 86 GHz \citep{had16}. Thus, $\alpha =0.7$ is adopted in the following.

\par Using Equation (4), we look for the best fit to the notion that the image in Figure 1 is consistent with approximate time averaged bilateral symmetry. From Section 3, in order for observed bilateral symmetry and Equation (4) to be justified, the analysis requires
\begin{enumerate}
\item Averaging or integrating over cross-sections of the jet and counter-jet. This is accomplished in Table 1 with two different approaches: the sum of peak ridge intensities and the flux density of 0.16 mas wide cross-sections.
\item Many cross-sections need to be compared in consort. This is achieved by sampling the data every 0.05 mas in Table 1.
\end{enumerate}
We do not assume a constant velocity, but a constant acceleration, $a$, the first order correction to a constant velocity.
Finding a solution is tantamount to finding the acceleration and initial velocity, $\beta_{0}$, of a flow starting at $z=\pm 0.4$ mas from the core, such that $\{R[\beta(z)]\}^{2.7}$ maps the counter-jet intensity at -z to the jet intensity at +z with the minimum residuals that is expressed in terms of the excess variance \citep{nan97}
\begin{eqnarray}
&&\Sigma_{\mathrm{rms}}^{2}=  \nonumber \\
&&\frac{1}{6}\sum_{i=1}^{6}\frac{\left(\{R[\beta(z)]\}^{2.7}I[z=-0.4-0.05(i-1)]\rm{mas}-I[z=0.4+0.05(i-1)\rm{mas}]\right)^2}{I[z=0.4+0.05(i-1)\rm{mas}]^{2}} \nonumber \\
&&-\frac{1}{6}\sum_{i=1}^{6}\frac{\left(\{R[\beta(z)]\}^{2.7}\sigma[z=-0.4-0.05(i-1)\rm{mas}]\right)^2 + \sigma[z=0.4+0.05(i-1)\rm{mas}]^2}{I[z=0.4+0.05(i-1)\rm{mas}]^{2}}\;,
\end{eqnarray}
where $I(z)$ is either the intensity from column (4) or the flux density from column (5) of Table 1. The relevant uncertainty, the second term of Equation (5), is for the difference, $\{R[\beta(z)]\}^{2.7}I(-z) -I(z)$, with errors propagated from the uncertainties, $\sigma(z)$, in columns (4) and (5) of Table 1. Empirically, it was found that the velocity distribution that was determined by minimizing the excess variance had a negligible dependence on the precise definition of uncertainty. The fit to the sum of the peak intensities of the ridges in the top panel of Figure 3 has the smallest $\Sigma_{\mathrm{rms}}^{2}$ for $a=145\rm{cm}-\rm{sec}^{-2}$, $\beta_{0}=0.281$.  The fit to the flux density of cross-sectional slices, in the lower left hand panel, has the smallest $\Sigma_{\mathrm{rms}}^{2}$ for $a=124\rm{cm}-\rm{sec}^{-2}$, $\beta_{0}=0.281$. The fact that they are similar is a nice consistency check. Most of the excess variance is generated by the excess flux density at z = 0.4 mas and noted in Table 1. This might be a result of core flux or a small flare in June 2011 aligned with the ejection time in the top panel of Figure 2. If the center of the cross-section is shifted by 0.005 mas (well below the resolution limits of the CC model and the array) to z = 0.405 mas, a strong CC, only 0.323 mas from the nucleus, does not contribute and we get the fit in the bottom right hand panel, $a=144\rm{cm}-\rm{sec}^{-2}$, $\beta_{0}=0.268$. This fit has an advantage over the peak ridge intensity fit since it utilizes the entire spatial cross-section and could be considered the most reliable fit. But, since all three fits are similar this distinction is of minimal consequence.
\begin{figure}[htp!]
\begin{center}
\includegraphics[width=170mm, angle=0]{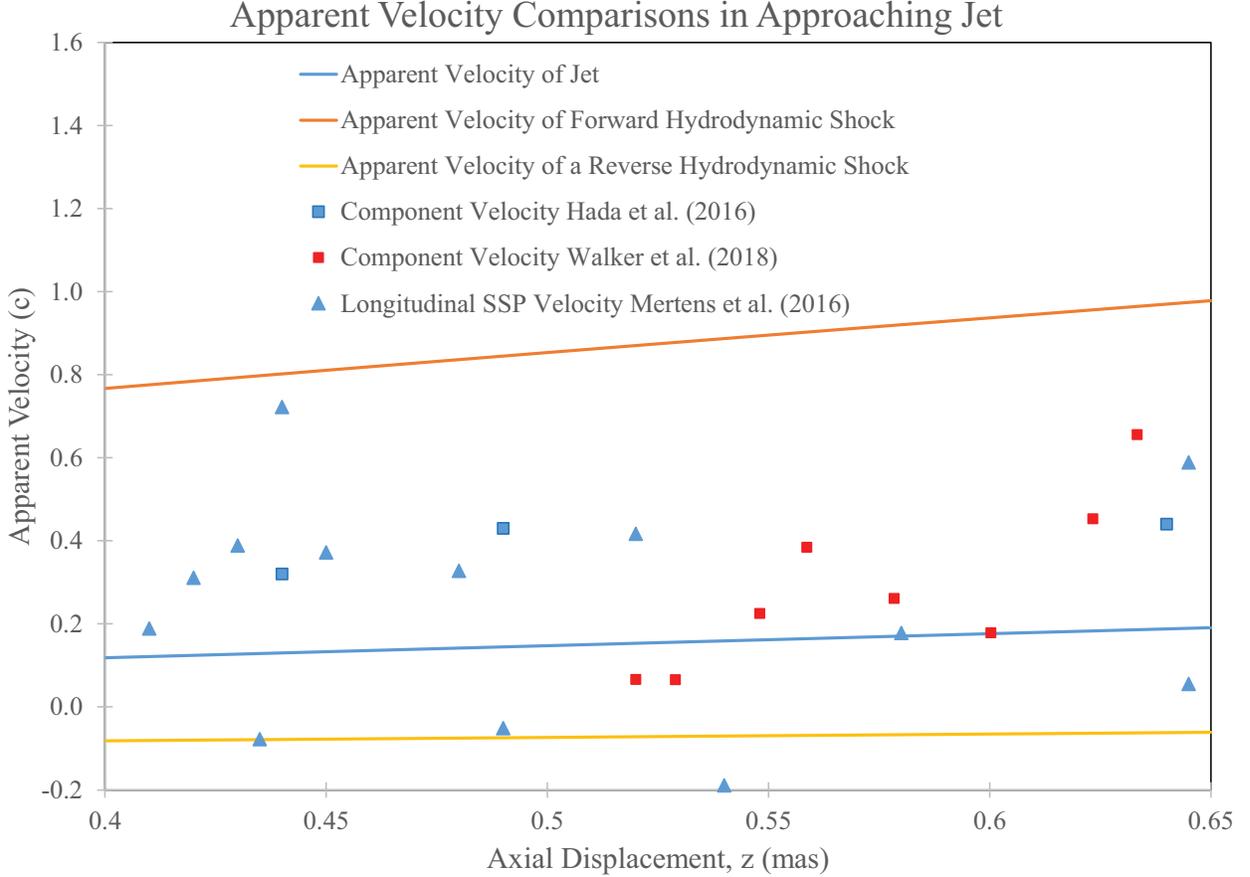}
\caption{A plot of the apparent velocity associated with the $\beta(z)$ from Figure 3. The apparent velocity of a forward and reverse hydrodynamic shock being advected by the flow are also plotted. The measured component velocities in the literature for $0.40 \rm{mas} < z < 0.65 \rm{mas}$ are scattered between these two boundaries. Thus, the moving components are consistent with being patterns from shock dissipation being advected by the flow velocity, $\beta(z)$. }
\end{center}
\end{figure}
\par The relativistic kinematics are rather benign compared to the highly relativistic motion inferred at $z\lesssim 900$ mas \citep{bir99,gir12}. The Lorentz factor, $\Gamma =1/\sqrt{1-\beta^{2}(z)}$, is 1.04 at $z=0.40$ mas and 1.08 at $z=0.65$ mas. The Doppler factor, $\delta = 1/[\Gamma (1- \beta(z)\cos{\theta})]$, is 1.30 at $z=0.40$ mas and 1.45 at $z=0.65$ mas.
\par The velocity estimates in this section refer only to the luminous portions of the jet which is concentrated on the bounding ridges as demonstrated in Table 1 and related discussions. The results cannot be applied to any non-radiating plasma. For example, there might be a central stream as has been detected at $z\gtrsim 2$ mas that flows faster than the ridges \citep{mer16}. If there is an under-luminous central stream at $z <0.65$ mas, it contributes minimally to the averaging of $\beta(z)$ in the $(x,y)$ plane in the fits shown in the bottom panels of Figure 3.
\section {Comparison to Component Motion Measurements} In order to compare these results to component motion found in the literature, we convert the intrinsic velocity, $\beta(z)$, in Figure 3, to the apparent velocity, $\beta^{\rm{apparent}}(z)$, using Equation (1). This is plotted in Figure 4. We also consider the notion of patterns due to simple hydrodynamic shocks in a relativistic fluid, $\beta_{\rm{shock}}=\sqrt{1/3}$ \citep{lig75}. The relativistic addition of a shock advected with the bulk flow, $\beta_{\rm{shock}}^{\rm{advected}}$ yields a net velocity \citep{lig75},
\begin{equation}
\beta_{\rm{shock}}^{\rm{advected}} = \frac{\beta(z)\pm \sqrt{1/3}}{1\pm \sqrt{1/3}\beta(z)}\;,
\end{equation}
where the plus sign indicates a shock launched in the flow direction (forward) and the minus sign is for a shock anti-directed to the flow (reverse). By combining the intrinsic velocity in Equation (6) with Equation (1), we plot the apparent forward and reverse shock velocities in Figure 4. We test the simple idea that the components detected in the flow are patterns from shocks superimposed on the bulk flow by including component velocities at $0.4 \rm{mas} < z < 0.65 \rm{mas}$ from the literature to Figure 4 \citep{had16,wal18}. We also include the``significant structural patterns" (SSP) longitudinal velocities from Figure 3 of \citep{mer16}. The data is consistent with the velocity determined in Section 4 being a bulk flow velocity and the measured components are dissipative shocks that are advected with the bulk flow.

\par In general, if the plasma is magnetized, there are three plasma wave modes, slow, intermediate and fast, in order of their propagation speed. For a magnetized plasma, the fast magneto-sonic speed will exceed the sound speed of hydrodynamic waves \citep{sti92}. The exact fast magneto-sonic speed depends on the unknown details of any significant magnetic field. In spite of not knowing the magnetic properties of the flow, including magnetization will expand the region between the forward shock and reverse shock apparent velocities in Figure 4. Thus, the effect of magnetization, does not change the interpretation of the component velocities (dissipative shocks that are advected by the flow) in Figure 4.

\par It should be emphasized that the constant acceleration approximation employed in this analysis is only valid over a small section of the jet. If one extrapolates to small distances, the three fits found in this section have zero velocity at $z\sim 0.08 \,- \,0.15$~mas which is not that accurate. Extrapolating outward, $\beta^{\rm{apparent}}(z)=6c$ occurs in the three fits in Figure 3 at $z=3.5\, - \,4.1$~mas. Apparent velocity of this magnitude has been estimated much farther out, $\sim$ 870 mas \citep{bir99,gir12,asa14}. Clearly, the acceleration must be smaller downstream of the fitted region in Figure 3. It is interesting that the velocity profile extrapolated farther out indicates that the flow becomes supersonic, $\beta(z) >\sqrt{1/3}$, at $z=1.31\, - \,1.36$ mas in the constant acceleration models of Figure 3. This is $\sim 1350$ M from the point of origin or $\sim 1.2$ lt-yrs.

\section{Conclusion}
In this study, the equations of special relativity are used to show that the high sensitivity image of M87 presented in Figure 1 is consistent with approximate time averaged intrinsic bilateral symmetry of the jet and counter-jet if appropriate spatial averaging is implemented. The solution, in Section 4, requires mild acceleration from 0.27c to 0.38c between 0.40 mas and 0.65 mas. The acceleration is $\approx 15\%$ of that from gravity at the Earth's surface. The velocity of the jet is $\approx 0.38$c (0.27 c) at a de-projected distance of 650 M (400 M) from the supermassive black hole. This is a very useful constraint for numerical models that address both the luminous 40$\mu$as diameter annulus detected by the Event Horizon Telescope and the base of jet. The jet is slow by relativistic and numerical model standards. The luminous outer ridges of the jet (the ``funnel wall jet") are already accelerated to 0.3c - 0.55c at 120-190 M from the black hole in 3-D numerical simulations \citep{dev05,kro05,mos16}.

The implications of the analysis in Sections 4 and 5, is that the velocity field derived in this study is likely a bulk flow for two reasons. Firstly, it does not depend on the motion of localized extreme surface brightness features, it uses the global surface brightness profile. Secondly, based on Figure 4, the distribution of the observed individual component velocity is well described hydrodynamic shock fronts moving on the background of the bulk flow velocity found in Section 4.

\begin{acknowledgments}
 This work benefited from the comments of a helpful referee that were much appreciated. I would like to thank R. Craig Walker for supplying the data for this paper. Not only that, but he generously spent the time to provide comments on multiple versions of this manuscript. The Very Long Baseline Array (VLBA) is an instrument of the National Radio Astronomy Observatory. The National Radio Astronomy Observatory is a facility of the National Science Foundation operated by Associated Universities, Inc.

\end{acknowledgments}

\end{document}